# Hydrogen (deuterium) dynamics and thermal stability in ion-irradiated platinum-hydride thin films synthesized at low temperature

S. S. Das[1,*], T. Ozawa[1,*], Y. Komatsu[2], R. Shimizu[2,†], T. Hitosugi[2,3] and K. Fukutani[1,4,*]

[1]*Institute of Industrial Science, The University of Tokyo, Komaba, Meguro-ku, Tokyo, 153-8505, Japan*

[2]*School of Materials and Chemical Technology, Institute of Science Tokyo, Ookayama, Meguro, Tokyo 152-8552, Japan.*

[3]*Department of Chemistry, The University of Tokyo, Hongo, Bunkyo, Tokyo 113-0033, Japan.*

[4]*Advanced Science Research Center, Japan Atomic Energy Agency, Shirakata, Tokai, Ibaraki 319-1195, Japan.*

*Present address:*

[†]*Institute for Molecular Science, National Institutes of Natural Science, Myodaiji, Okazaki, Aichi 444-8585, Japan.*

**Abstract:** The interaction of hydrogen (H) and deuterium (D) with transition metals plays a central role in heterogeneous catalysis reactions and hydrogen-related technologies. While H–Pt surface interactions have been extensively studied, direct experimental investigations of hydrogen incorporation and transport in Pt remain limited due to its intrinsically low solubility. Here, we present a comprehensive study of H(D) incorporation and desorption dynamics in metastable $PtH(D)_x$ thin films prepared using low-energy ion irradiation, enabling controlled hydrogen loading far above equilibrium concentrations. Nuclear reaction analysis (NRA) reveals a nonuniform hydrogen depth profile with two distinct hydrogen accumulation regions: one in the subsurface and the other near the film-substrate interface. Thermal desorption spectroscopy (TDS) exhibits two distinct desorption peaks near 190 and 230 K, consistent with hydrogen release from these two sites. Resistance relaxation measurements, analyzed within a two-parallel-channel conduction model indicate two different relaxation kinetics for subsurface and near-interface hydrogen. Arrhenius analysis reveals two thermally activated processes for

[*]Corresponding authors: ssdas@iis.u-tokyo.ac.jp (S. S. Das), t-ozawa@iis.u-tokyo.ac.jp (T. Ozawa) and fukutani@iis.u-tokyo.ac.jp (K. Fukutani)

$PtH_x$ with an average hydrogen concentration of $x = 0.15$, with activation energies of 130 ± 18 meV (subsurface) and 164 ± 26 meV (near interface). In the thermally activated regime (>140 K), D exhibits systematically slower relaxation rates with activation energies of 117 ± 8 and 121 ± 7 meV for $PtD_x$ prepared under the same implantation dose. Within the experimental uncertainty, the activation barriers remain comparable, while the prefactors reduce significantly for D, indicating isotope-dependent attempt frequencies and zero-point energy effects. TDS simulations based on the Polanyi-Wigner formalism qualitatively reproduce the experimental desorption spectra by resolving distinct subsurface and near-interface contributions, in agreement with the NRA depth profile. These findings provide insight into hydrogen kinetics in $PtH_x$ relevant to Pt-based catalysis, sensing, and hydrogen–metal interactions.

**Graphical Abstract**

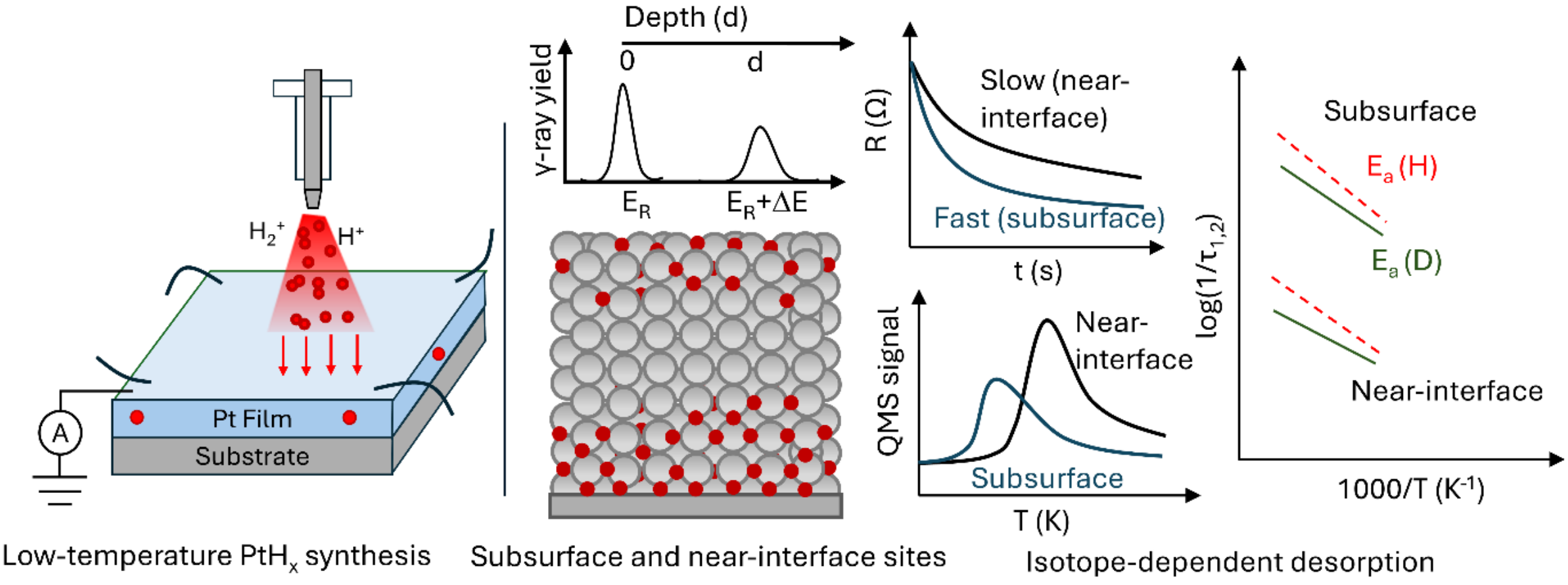

Low-temperature synthesis of $PtH_x$ thin films induces subsurface and interface hydrogen sites that govern double-exponential relaxation and desorption kinetics.

Keywords: Platinum hydride, Nuclear reaction analysis, Thermal desorption spectroscopy, Hydrogen desorption, Kinetic simulation, Time-dependent relaxation.

## I. INTRODUCTION

Hydrogen is the lightest and most mobile interstitial species, capable of diffusing rapidly through the metal lattice. Its interaction with transition metals is a major focus of research across various fields such as surface science, catalysis, hydrogen storage, sensing, and energy conversion technologies [1–7]. Platinum (Pt), owing to its exceptional catalytic activity and favorable hydrogen binding energetics, is widely used in hydrogen evolution and oxidation reactions [8,9]. Previous studies on Pt-H systems have extensively investigated hydrogen adsorption/desorption, surface reconstruction, spill over, and catalytic reaction pathways on low-indexed Pt surfaces [10–12]. In contrast, the microscopic mechanisms governing hydrogen absorption and desorption within Pt remain far less understood and are still actively debated [12,13].

Besides this, another significant challenge is the formation of a stable platinum hydride. Under ambient conditions, Pt does not form a hydride, and conventional gas-loading methods fail to achieve appreciable hydrogen concentrations [12]. However, recent high-pressure studies have reported the formation of stable platinum hydride $PtH_x$($x \sim 1$) at elevated pressure (> 27 GPa) and room temperature [14–17]. Depending on pressure, $PtH_x$ exhibits three reported phases: PtH-I, PtH-II, and a mixed PtH I/II phase [14]. At room temperature, the mixed phase persists up to ~ 42 GPa, above which the PtH-II single phase forms. Although the crystal structure of these two phases are not fully resolved, it is suggested that H atoms preferentially occupy tetrahedral (T) interstitial sites in the PtH-I phase and octahedral (O) sites in the PtH-II phase [14,17]. Interestingly, the PtH-II phase exhibits superconductivity with a critical temperature of 6.7 K, nearly three orders of magnitude larger than pure Pt. These high-pressure PtH-I and PtH-II phases show that platinum can accommodate hydrogen in interstitial hydride states under extreme equilibrium conditions. Although the hydrogen states in the present study are not direct equivalents of these bulk high-pressure phases, they represent metastable hydrogen states stabilized in Pt thin films, providing complementary insight into hydrogen accommodation and transport in Pt. Apart from hydride formation under extreme conditions, experimental studies on bulk hydrogen diffusion in Pt are scarce due to its extremely low solubility [12]. Early permeation measurements on high purity single-crystal Pt revealed H solubilities in the range of $10^{-5}$ to $10^{-11}$ (H/Pt), depending on temperature, reflecting a large positive enthalpy of solution (11 kcal/mol.K). Despite this poor solubility, the intrinsic bulk diffusion coefficient follows Arrhenius behaviour with an activation energy of 0.26 eV and a preexponential factor of the order of $10^{-3}$ $cm^2/s$ in the temperature range of approximately 615–925 K. Measurements on hydrogen and deuterium showed only a small isotope effect in their diffusion coefficients ($D_H/D_D \sim 1.16$), implying that hydrogen diffuses only marginally faster than deuterium. In another study [13], high-temperature permeation measurements in Pt have shown that the bulk diffusion coefficient follows Arrhenius behaviour with an activation energy of 0.27 eV over a wide temperature range 558-936 ℃. It also highlights the low hydrogen solubility in Pt, underscoring the difficulty of directly probing hydrogen dynamics in bulk Pt. Despite several works, an experimental study on the growth of $PtH_x$ under a vacuum condition and its thermal stability has not been achieved so far. Ion-implantation, plasma-assisted hydrogenation, and low-energy ion irradiation techniques have shown significant promise to generate metastable $PtH_x$ with significantly enhanced H solubility [18,19]. These metastable hydrides provide unique opportunities for quantitative analysis of hydrogen concentration profiles, diffusion kinetics, isotope effects, and quantum transport phenomena that are inaccessible through conventional hydrogen loading techniques.

In this context, a Pt thin film offers a unique platform to correlate hydrogen concentration, depth distribution, and kinetic behavior using advanced experimental methods involving nuclear reaction analysis (NRA), thermal desorption spectroscopy (TDS), and temperature-dependent resistance relaxation measurements. Combining these techniques provides a comprehensive picture of hydrogen dynamics that are not accessible through surface-only or bulk-only approaches. Here, we report a detailed study of hydrogen (H) and deuterium (D) diffusion/desorption dynamics in $PtH(D)_x$ thin films grown at low temperature using a low-energy ion irradiation technique. NRA depth profiles reveal non-uniform hydrogen distribution, with accumulation in the subsurface region and at the film-substrate near-interface region. Temperature-dependent resistance relaxation systematically distinguishes the hydrogen/deuterium dynamics in the Pt film. By combining depth profiling, desorption analysis, and resistance relaxation dynamics, this work provides new microscopic insight into H(D) behaviour in Pt thin films and offers new insights for designing Pt-based catalytic and hydrogen sensing materials.

## II. EXPERIMENTAL METHODS

Polycrystalline Pt thin films with a nominal thickness of 20 nm were deposited onto glass substrates with a size of 10 × 10 mm$^2$ at room temperature using a conventional magnetron sputtering system. The active area of the sample used for implantation and in situ measurements was approximately 8 × 8 mm$^2$. The base pressure of the chamber was 2 × 10$^{-4}$ Pa, and sputtering was carried out at an Ar pressure of 1.0 Pa. The Pt films were deposited at a rate of 10 nm/min by applying a DC sputtering power of 30 W. The structural integrity and the phase purity of the films were confirmed post-growth using two-dimensional X-ray diffraction (2D-XRD) mapping using a Rigaku R-AXIS RAPID II system (See supplementary Fig. S1), which verified polycrystalline nature with a characteristic (111) texture.

Hydrogen and deuterium ions were generated from $H_2$ and $D_2$ gases, respectively, using a Varian ion gun and subsequently implanted on to Pt films at an energy of 500 eV and a sample temperature of 25 K, forming metastable $PtH(D)_X$. Since the main ion species generated by the ion gun is molecular hydrogen/deuterium ions, the energy per H(D) atom is expected to be 250 eV. No additional surface treatment was performed prior to H/D loading. The irradiation dose was systematically varied in the range of 0–156 mC/cm$^2$ to control the incorporated H(D) concentration. The irradiation configuration and the ion energy used for H and D loading were identical. Stopping and Range of Ions in Matter (SRIM) simulations were also performed to estimate the implantation profiles of 250 eV H(D) atoms in the Pt film. The simulations indicate negligible Pt sputtering under the present irradiation conditions. NRA, resistance relaxation, and TDS measurements were performed for both hydrogen- and deuterium-implanted Pt films at selected implantation dose of 4.7 mC/cm$^2$, which provides a moderate H(D) concentration before strong saturation.

Hydrogen dynamics were studied using resistance-relaxation experiments performed in the temperature range 5–200 K in a UHV chamber with a base pressure of 5 × 10$^{-7}$ Pa. The deposition and relaxation measurements were performed in separate vacuum systems. Prior to the measurements, several H(D) loading/unloading cycles were performed to reduce weakly adsorbed surface species. Therefore, although minor native oxide or strongly bound adsorbates cannot be completely excluded, their influence in the observed behaviour is expected to be limited. No capping layer was used in the present study. The hydrogenation, resistance-relaxation, and TDS measurements were performed sequentially in the same UHV chamber after implantation under identical in-situ measurement geometry. In these experiments, the $PtH(D)_x$ film was rapidly brought to the desired temperature immediately after the implantation, and the time evolution of the electrical resistance was continuously recorded, similar to that in Ref. [20,21]. Electrical resistance was measured in a fixed four-probe configuration, with contacts placed at the four corners of a square-shaped film. A Keithley 2000 digital multimeter was used in resistance mode with a probe current of 1 mA, and the resistance output was recorded continuously during the relaxation process. Hydrogen/deuterium desorption from the $PtH(D)_x$ film was examined by TDS experiment using a quadrupole mass spectrometer (QMS, Pfeiffer Vacuum) while heating the sample at a constant rate of ~ 0.4 K/s. The desorbed $H_2$ and $D_2$ signals were monitored at mass-to-charge ratios of 2 and 4, respectively. The total hydrogen content was estimated from the integrated area of the TDS spectrum after the background subtraction. The obtained TDS peak areas were mainly used for relative comparison of the hydrogen desorption, while the absolute hydrogen concentration was estimated from the NRA measurements.

Electrical resistance was monitored in real time throughout the entire cycle of hydrogenation, relaxation, and dehydrogenation. The hydrogen depth distribution in the Pt film was investigated by resonant $^{1}H(^{15}N, \alpha\gamma)^{12}C$ Nuclear Reaction Analysis (NRA) experiment in a separate UHV chamber (base pressure: $3 \times 10^{-7}$ Pa) at the Micro Analysis Laboratory, Tandem accelerator (MALT) facility of the University of Tokyo [22]. A $^{15}N^{2+}$ ion beam with a size of 3 mm in diameter and a current of about 10 nA was incident to the sample in the surface-normal direction, and the γ-rays produced from the NRA reaction were detected using a $Bi_4Ge_3O_{12}$ scintillation detector placed behind the sample. The $^{15}N$ beam energy was scanned across the resonance energy of 6.385 MeV to obtain hydrogen depth profiles with a near-surface depth resolution of ~2–10 nm. The H concentration was evaluated with the γ-detection efficiency of the setup calibrated by a Kapton foil with a known H concentration [22]. NRA measurements were performed at selected temperatures and irradiation doses for the $PtH_x$ films.

## III. RESULTS AND DISCUSSION

### A. H-depth distribution in $PtH_x$

The quantitative information on hydrogen depth distribution in the $PtH_x$ film was obtained from NRA experimental data collected at 25 K. Here, we define $x_{Ave}$ as the average hydrogen concentration in Pt thin films estimated from the integrated NRA profile, whereas the actual hydrogen distribution within the film is nonuniform, as discussed below. Figure 1

(a) shows the normalized γ-yield as a function of the $^{15}$N-ion energy for different hydrogen-ion implantation doses ($H_{imp}$) in the range 0–156 mC/cm$^2$. The NRA profile shows two peaks located at depths of around 5 and 13 nm from the surface. SRIM simulations for 250 eV H and D ions in Pt (see Supplementary Fig. S2) predict implantation ranges of approximately 3.7 and 4.4 nm, respectively. This suggests that the directly implanted H(D) ions are initially confined within the near-surface region of the Pt(20 nm) film. Thus, hydrogen accumulation in the deeper region (~13 nm) cannot be explained only on the basis of ballistic implantation; instead, it suggests the presence of stable trapping sites with a limited post-implantation redistribution within the Pt film. These two features in the NRA profile imply two macroscopically stable hydrogen sites in the Pt film, i.e. subsurface and film-substrate near-interface. With increasing $H_{imp}$ dose, both peaks grow in intensity and approach saturation for $H_{imp} \geq 18.8$ mC/cm$^2$, while their width changes only marginally. The average hydrogen concentration extracted from the integrated NRA yield increases nearly linearly with $H_{imp}$ before saturation at x = 0.37, as shown in Fig. 1(b) (left axis). The in-situ electric resistance recorded during the hydrogen implantations and NRA measurements (Fig.1(b), right axis) shows an increase with $H_{imp}$ before saturation. While a linear fit captures the trend only over an intermediate concentration range (x = 0.1–0.4), it fails to describe the data across the full range. In contrast, a parabolic fit provides a significantly better representation of the entire range (x = 0.0–0.4). This nonlinear behaviour in resistance change suggests that the resistance response cannot be attributed to a single hydrogen-induced mechanism, but instead reflects the combined contribution of multiple hydrogen-related effects such as changes in carrier scattering, lattice distortion, and hydrogen–hydrogen interactions. The nominal content of implanted hydrogen in the film was compared with the NRA-estimated value to estimate the hydrogen retention fraction (see Supplementary Table S1). The retained hydrogen fraction shows a decrease from approximately 61% at 1.56 mC/cm² to about 3 % at 156.25 mC/cm² with the increase of implantation dose, while the average hydrogen concentration approaches saturation around $x = 0.4$. This behavior suggests that only a portion of the implanted hydrogen atoms is retained in the film, and that the remainder is likely lost through ion backscattering/reflection, surface recombination, and immediate desorption during implantation. These combined NRA and resistance results clearly demonstrate controlled hydrogen incorporation in Pt and a relationship between the H concentration and electric transport.

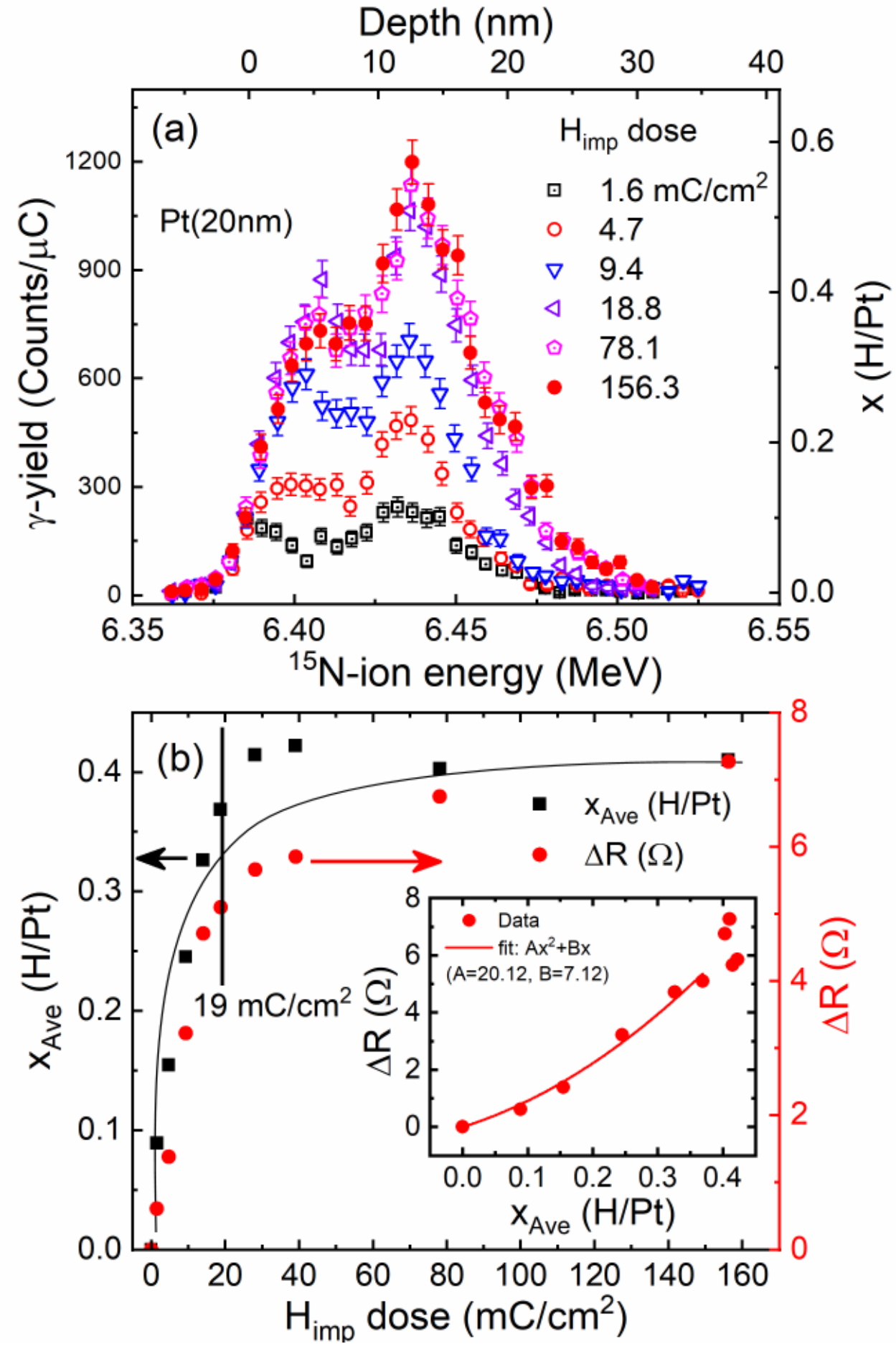

**Fig. 1** (a) Hydrogen-depth profiles of the $PtH_x$ film obtained from the NRA measurement after 500 eV hydrogen-ion implantation at 25 K. The γ-yield, normalized to the incident $^{15}N$ beam dose, is plotted on the left y-axis as a function of $^{15}N$ beam energy. The corresponding local hydrogen concentration (x) is shown on the right y-axis, with the depth on the top x-axis. (b) Average hydrogen concentration ($x_{Ave}$) extracted (left y-axis) from the NRA-depth profile data and change in resistance ΔR (right y-axis) due to hydrogenation plotted as a function of the implantation dose. The solid curve in (b) is a guide to eyes. The inset of panel (b) shows ΔR as a function of hydrogen concentration x. The solid curve represents a parabolic fit, $\Delta R = 20.12\,x^2 + 7.12\,x$. The clustered high-concentration points near x ≈ 0.4, where the resistance response tends to saturate, were excluded from the fitting.

### B. Thermal Stability and Desorption Behavior of Metastable $PtH_x$

The thermal stability of the metastable $PtH_x$ films was examined through in-situ NRA, resistance measurements, and TDS during heating. Figure 2(a1, a2) shows the NRA profiles measured at different temperatures for doses 4.7 and 15.6 mC/cm$^2$. After the sample was stabilized at the desired temperature, the $^{15}$N beam was applied, and NRA measurements were conducted. The sample was subsequently heated to the next selected temperature for NRA measurement. Thus, the $^{15}$N beam was applied intermittently rather than continuously during the warming process. Both figures show a gradual decrease in H concentration as temperature rises. Figure 2(b1, b2) presents the temperature dependence of the γ-yields measured at depths of approximately 6 nm and 12 nm, corresponding to subsurface and near-interface regions, respectively, for hydrogen implantation doses of 4.7 and 15.6 mC/cm$^2$. For the 4.7 mC/cm$^2$-implanted sample, the hydrogen depth profiles show a prominent interface peak accompanied by a comparatively weaker subsurface contribution. This indicates that at lower doses, hydrogen preferentially accumulates at the near-interface region, with limited stabilization in the subsurface region. In contrast, for the 15.6 mC/cm$^2$-implanted sample, hydrogen is observed in both the subsurface and near-interface regions. As the temperature increases, the γ-yield associated with the subsurface region decreases more rapidly than that of the near-interface region (Fig. 2(b2)), indicating a faster desorption of the subsurface hydrogen. Both the subsurface and near-interface γ-yields approach their respective background levels near 180

K, which is consistent with the near-complete removal of hydrogen from both regions. On the other hand, our TDS data (Fig. 3c) shows hydrogen desorption signal existing up to 350-450 K. The apparent difference between the temperature ranges observed in NRA and TDS may be attributed to an experimental artifact arising from ion-induced reduction during the NRA measurement. It should be noted that, because the NRA measurements were conducted at selected temperatures rather than continuously during the full temperature ramp, possible ion-beam-induced reduction is minimized, although it may still partly cause signal decrease. The persistence of the near-interface NRA signal up to higher temperature suggests that near-interface hydrogen has higher apparent thermal stability and may experience stronger binding than subsurface hydrogen.

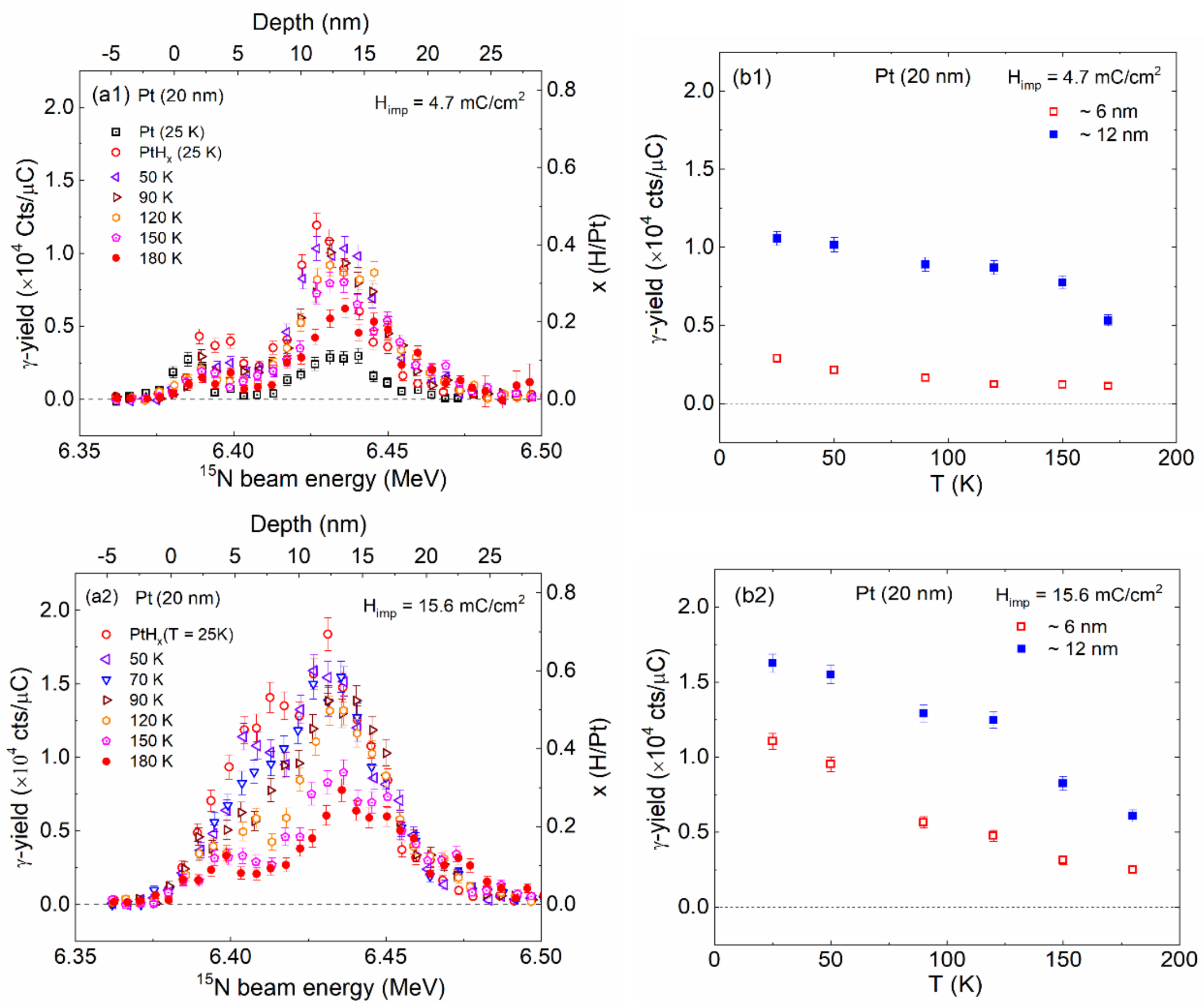

**Fig. 2** (a1, a2) γ-yield normalized to the incident $^{15}$N beam dose as a function of $^{15}$N beam energy (left y-axis) measured at different temperatures for Pt(20 nm) films hydrogenated with implantation doses of $H_{imp}$ = 4.7 mC/cm$^2$ (a1) and 15.6 mC/cm$^2$ (a2). Corresponding depth from the surface is shown on the top x-axis. (b1, b2) Temperature-dependent γ-yield extracted at depths 6 nm (red open symbol) and 12 nm (blue filled symbol) for $H_{imp}$ = 4.7 mC/cm$^2$ (b1) and 15.6 mC/cm$^2$ (b2).

The electrical resistance was continuously monitored in real time during both NRA and TDS runs to correlate electric transport with hydrogen loading and release. The Pt film was hydrogenated up to a dose of 15.6 mC/cm$^2$ at 25 K, corresponding to an average hydrogen concentration of $x_{Ave}$ =0.34. Figure 3(a) shows the temperature dependence of the NRA γ-yield measured at a fixed $^{15}$N ion energy of 6.429 MeV, whose probing depth is 12 nm from the surface, corresponding to the near-interface peak in the H-depth profile data shown in Fig. 1(a).

The NRA γ-yield shown in Fig. 3(a) decreases gradually with increasing temperature, with a more pronounced reduction above ~90 K and a sharp decrease above ~140 K. The signal remains low at higher temperatures up to the measured range of ~490 K, indicating near-complete hydrogen removal from the probed near-interface region. A similar temperature dependence is also observed in the in-situ resistance data shown in Fig. 3(b), which were recorded simultaneously with the NRA measurements in Fig. 3(a). The resistance change (ΔR) with respect to the resistance at 25 K before hydrogenation increases to approximately 6 Ω by hydrogenation at 25 K. Upon warming the sample, ΔR begins to decrease above 90 K, indicating the onset of hydrogen release from the film. The sample was heated up to 490 K, and the pure metallic state of Pt is recovered, which is reflected in the linear temperature dependence of ΔR recorded during the subsequent cooling.

The desorption behaviour was further examined using TDS experiments. Figure 3(c) shows the $H_2$ desorption spectrum for doses of 3.1, 4.7, and 6.3 mC/cm$^2$, recorded while warming the sample up to 490 K. At low doses (3.1 and 4.7 mC/cm$^2$), the TDS spectra consist of two overlapping broad peaks, one at 230 K and the other around 350 K. With increasing dose (≥ 6.3 mC/cm$^2$), an additional low-temperature peak near ~190 K emerges, while the two desorption components near ~190 and ~230 K become more clearly distinguishable, suggesting the presence of hydrogen populations with different binding strengths and thermal stabilities. This aligns with the two-peak feature observed in the H depth profile (Fig. 1(a)), corresponding to subsurface and near-interface hydrogen accumulation. As seen in Fig. 2(b1) and (b2), the near-interface component desorbs at higher temperature than the subsurface component. This suggests that the hydrogen in the near-interface region is more stable and has higher binding energy with respect to the subsurface region. Therefore, the observed low- and high-temperature TDS peaks are assigned to subsurface and near-interface hydrogen, respectively.

The desorption spectra of $PtD_x$ film prepared under the same implantation dose, shown in Fig. S3, also confirms the two-peak feature.

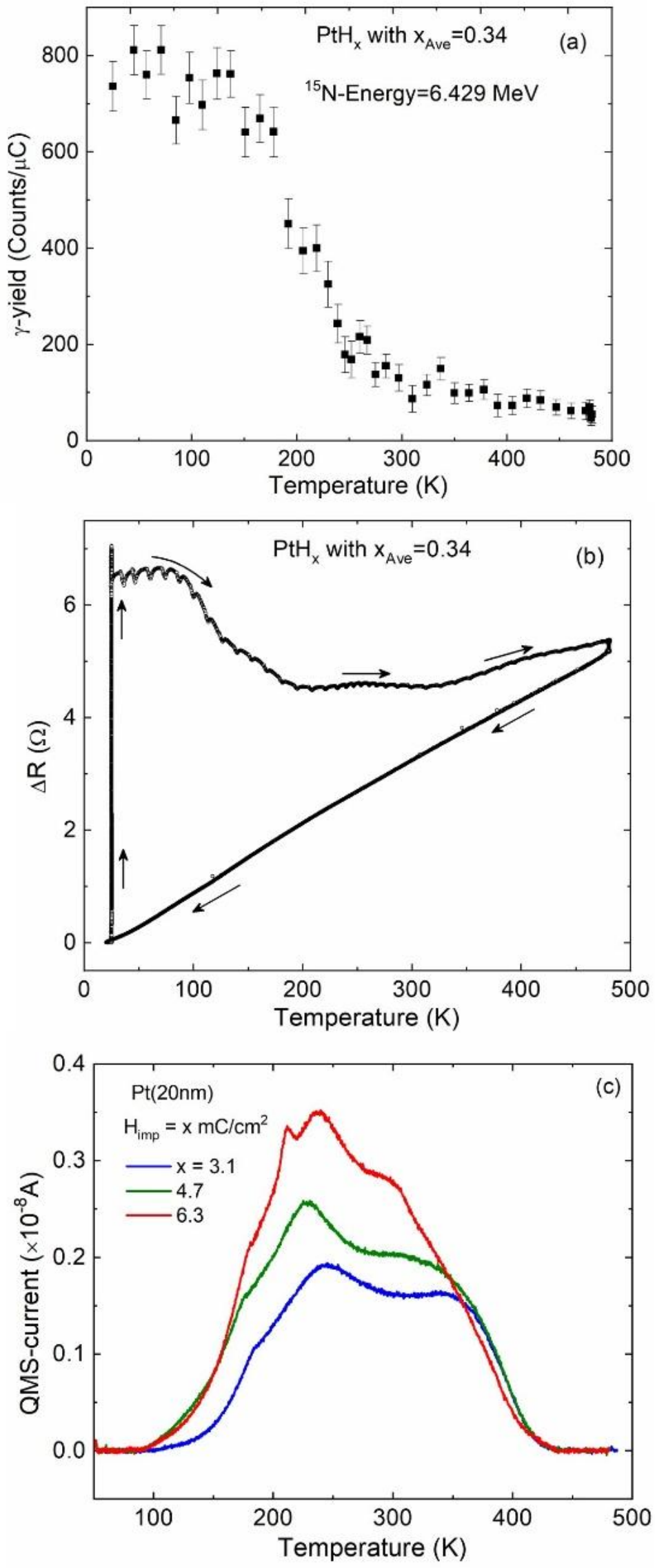

**Fig. 3** (a) Temperature dependence of the normalized NRA γ-yield at a $^{15}$N ion energy of 6.429 MeV for the $PtH_x$ film with $x_{Ave} = 0.34$, hydrogenated by 500 eV hydrogen ions at 25 K. (b) Temperature dependence of the resistance change ΔR with respect to the resistance at 25 K before hydrogenation recorded in situ while heating the sample up to 490 K. The cooling curve reflects the recovery of the pure metallic state of Pt. (c) TDS spectra of $PtH_x$ films as a function of temperature for various implantation doses of 3.1, 4.7, and 6.3 mC/cm$^2$, measured in a separate chamber. The TDS data reveal the onset of hydrogen desorption at approximately 100 K, with two prominent desorption peaks around 190 K and 230 K.

### C. Hydrogen Dynamics in $PtH_x$

Hydrogen desorption dynamics in the $PtH_x$ films were investigated through the time evolution of resistance at selected target temperatures immediately after hydrogenation at 25 K. Figure 4 shows the normalized resistance $R/R_T$ as a function of time in the temperature range 60–180 K, where $R_T$ represents the resistance at the beginning of each relaxation measurement. Below 70 K, the resistance remains nearly unchanged over the entire measurement time. At 100 K, the resistance begins to decrease gradually, and relaxation becomes faster as the temperature increases above 140 K. Since hydrogen desorption from $PtH_x$ film onsets above 100 K, as observed in the TDS spectra in Fig. 3(c), the observed resistance decrease is attributed to hydrogen release from the sample.

The resistance relaxation data for the $PtH_x$ film with $x_{Ave} = 0.15$ in Fig. 4 exhibit double-exponential features. The presence of two exponential terms suggests two distinct resistance channels with different H release rates in $PtH_x$. This interpretation is consistent with the earlier observation of two peaks in TDS (Fig. 3(c)) and NRA (Fig. 1(a)) measurements, which are attributed to subsurface and near-interface hydrogen populations. To verify the physical relevance of two resistance relaxation channels in the $PtH_x$ sample, we employ a simple illustrative model by assuming the film consists of two parallel conduction channels and correlate the relaxation and thermal desorption. Each channel is assumed to relax independently from a high resistive metal-hydride state to a low resistive pure metallic state with first-order kinetics, with respective activation energies and prefactors. It should be noted that this two-channel model in the present analysis is a simplified phenomenological model. Although it is consistent with the NRA-depth profile for the existence of two hydrogen populations in subsurface and near-interface regions, several other factors such as distribution of trapping sites, grain-boundary effects, or a range of diffusion barriers may also contribute to the bi-exponential relaxation behavior. Therefore, the extracted fast and slow components here should be regarded as effective relaxation processes rather than independent conduction channels.

The time dependence of the resistance of these two channels can be described as

$$r_1 = r_{A0} + ae^{\left(-\frac{t}{\tau_1}\right)}, \qquad (1)$$

$$r_2 = r_{B0} + be^{\left(-\frac{t}{\tau_2}\right)}. \qquad (2)$$

Here, $r_{A0}$ and $r_{B0}$ represent the initial resistances of the two respective regions in the metallic phase, while $a$ and $b$ denote the amplitudes of the hydrogen-induced resistance change, and $\tau_1$ and $\tau_2$ are the characteristic time constants. The quantities $r_{A0}$ and $r_{B0}$ become identical when the thickness of the two regions become equal. For simplicity, we assume that the subsurface and near-interface regions have comparable effective thickness and similar intrinsic resistivity in the hydrogen-free state. Under this approximation, the corresponding baseline resistances become equal, i.e., $r_{A0} = r_{B0} = r_0$. Accordingly, the total resistance $R_{Total}$ of the $PtH_x$ film can be written as,

$$R_{Total} = \left(\frac{r_1 r_2}{r_1 + r_2}\right) = \frac{\left(r_0 + ae^{\left(-\frac{t}{\tau_1}\right)}\right)\left(r_0 + be^{\left(-\frac{t}{\tau_2}\right)}\right)}{2r_0 + ae^{\left(-\frac{t}{\tau_1}\right)} + be^{\left(-\frac{t}{\tau_2}\right)}}. \qquad (3)$$

This shows that the expression for $R_{Total}$ contains two intrinsic time scales, leading to an apparent bi-exponential relaxation, where the early-time response ($\tau_1$) is dominated by the faster channel corresponding to subsurface-H release and the long-time response ($\tau_2$) is governed by the near-interface-H release. However, the near-interface component should not be regarded as entirely independent, since hydrogen released from the near-interface region must pass through the overlying Pt layer and subsurface region before getting desorbed from the surface. Therefore, the subsurface and near-interface components represent coupled hydrogen-release pathways associated with different depth regions. The relaxation data for the $PtH_x$ film with $x_{Ave} = 0.34$ and the $PtD_x$ film prepared under the same implantation dose as the $PtH_x$ film with $x_{Ave} = 0.15$ were analyzed using the same formalism, and the corresponding results are presented in Figures S4 and S5.

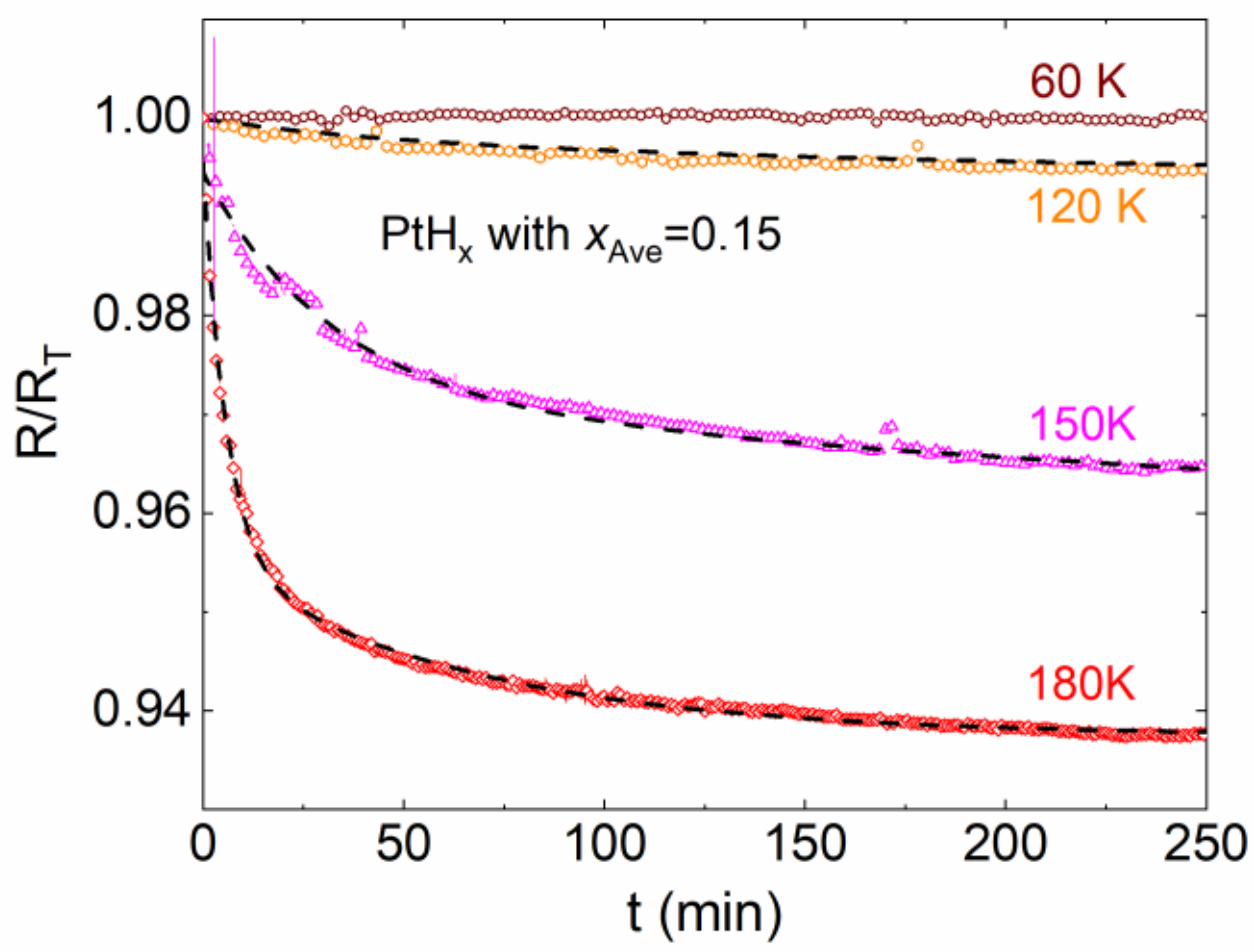

**Fig. 4** Time evolution of the normalized resistance $R/R_T$ of the $PtH_x$ film with $x_{Ave} = 0.15$ measured at fixed temperatures of 60, 120, 150, and 180 K, where $R_T$ denotes the resistance at the measurement temperature immediately before the onset of relaxation. The resistance relaxation is negligible at 60 K, and becomes progressively more pronounced with increasing temperature in the range 120–180 K. The dashed curves represent fits to the experimental data using equation (3).

The characteristic time constants $\tau_1$ and $\tau_2$ obtained from the fitting of the relaxation curves using eq. (3) were further analyzed using the Arrhenius plot to derive the corresponding activation energies. The inverse of the time constants, $1/\tau_1$ and $1/\tau_2$, represent the H(D) release rates associated with the subsurface and near-interface relaxation components, respectively. Figure 5 shows the temperature dependence of the $1/\tau_1$ and $1/\tau_2$ for $PtH_x$ film with $x_{Ave} = 0.15$ and the corresponding $PtD_x$ film prepared under the same implantation dose. The relaxation rate exhibits a clear exponential increase above 140 K, consistent with thermally activated desorption behaviour that can be described by the Arrhenius relation $1/\tau = 1/\tau_0 \exp(-E_a/k_BT)$. For $PtH_x$ film with $x_{Ave} = 0.15$, the subsurface and near-interface components yield activation energies ($E_a$) of 130 ± 18 and 164 ± 26 meV, respectively, while the corresponding prefactors (14.2 ± 3.6 and 11.9 ± 6.04 $s^{-1}$) remain comparable within the experimental uncertainty. The TDS peak temperatures and extracted kinetic parameters for H and D are summarized in Table 1. This suggests that the two relaxation processes operate on similar kinetic energy scales, despite originating from different regions and exhibiting distinct relaxation time constants. The larger activation energy for the near-interface component is consistent with the NRA results, which indicate higher apparent thermal stability of hydrogen near the interface than in the subsurface region. As the hydrogen concentration increases, the Arrhenius analysis for $PtH_x$ with $x_{Ave} = 0.34$ (Fig. S6) reveals a systematic decrease in both the activation energy and prefactor, consistent with concentration-induced modifications to the diffusion mechanism or enhanced trapping interactions. Similarly, in the same thermally activated regime (> 140 K), the relaxation rates for deuterium were systematically investigated. These rates were found to be lower than those for hydrogen over the entire temperature range. Arrhenius fits yield activation energies of 117 ± 8 meV for the subsurface component and 121 ± 7 meV for the near-interface component in the corresponding $PtD_x$ film prepared under the same implantation dose as the $PtH_x$ of $x_{Ave} = 0.15$. Within the fitting uncertainty, these $E_a$ values are comparable or slightly lower than those for hydrogen, implying the influence of the mass-dependent zero-point energy. In contrast, the corresponding prefactors for D are largely reduced, with values of 3.4 ± 1.8 (subsurface) and 0.39 ± 1.6 $s^{-1}$(near-interface). Large uncertainties in the latter reflect the limited sensitivity of the Arrhenius intercept for very slow relaxation processes. Nevertheless, the prefactors for D remain much smaller than those of H, reflecting the lower attempt frequencies of the heavier isotope. This behaviour points to an apparent isotope effect in the activation barriers derived from Arrhenius analysis, indicating that the effective activation barriers are influenced by zero-point energy contributions rather than purely classical, mass-controlled barrier crossing.

In contrast to the Arrhenius relation above 140 K, in the low-temperature regime below 140 K, the relaxation rates for both H and D deviate from the Arrhenius relation and become nearly temperature-independent. This weak temperature-dependence suggests that the relaxation process is not governed solely by the classical thermally activated barrier crossing in this regime. Instead, quantum effects may contribute to the deviation from Arrhenius behavior at low temperature [23,24]. Similar low-temperature deviations from Arrhenius behaviour have been discussed in hydrogen diffusion studies in transition metals, where lattice- and electron-coupled hydrogen motions become important in tunneling at low temperature [1,20,25–29].

In the present $PtH(D)_x$ film, the low-temperature relaxation behaviour is similar for H and D. This observation suggests that the mechanism cannot be simply attributed to tunneling of a single hydrogen atom, which would generally be expected to exhibit a pronounced isotope effect. Instead, the relaxation may involve coupled hydrogen–lattice motion or collective rearrangement of the local Pt–H(D) environment, which could reduce the apparent isotope dependence. In Pd films, phonon-assisted resonant tunnelling of proton between the interstitial sites has been observed, while isotope-independent transport of hydrogen at lower temperature has been discussed in terms of correlated hydrogen motion and collective effects [20]. At present, however, the microscopic origin of the low-temperature relaxation remains unresolved. Therefore, we refer to the low-temperature regime as a low activation relaxation regime, while noting that further theoretical and computational studies would be required to clarify the underlying mechanism.

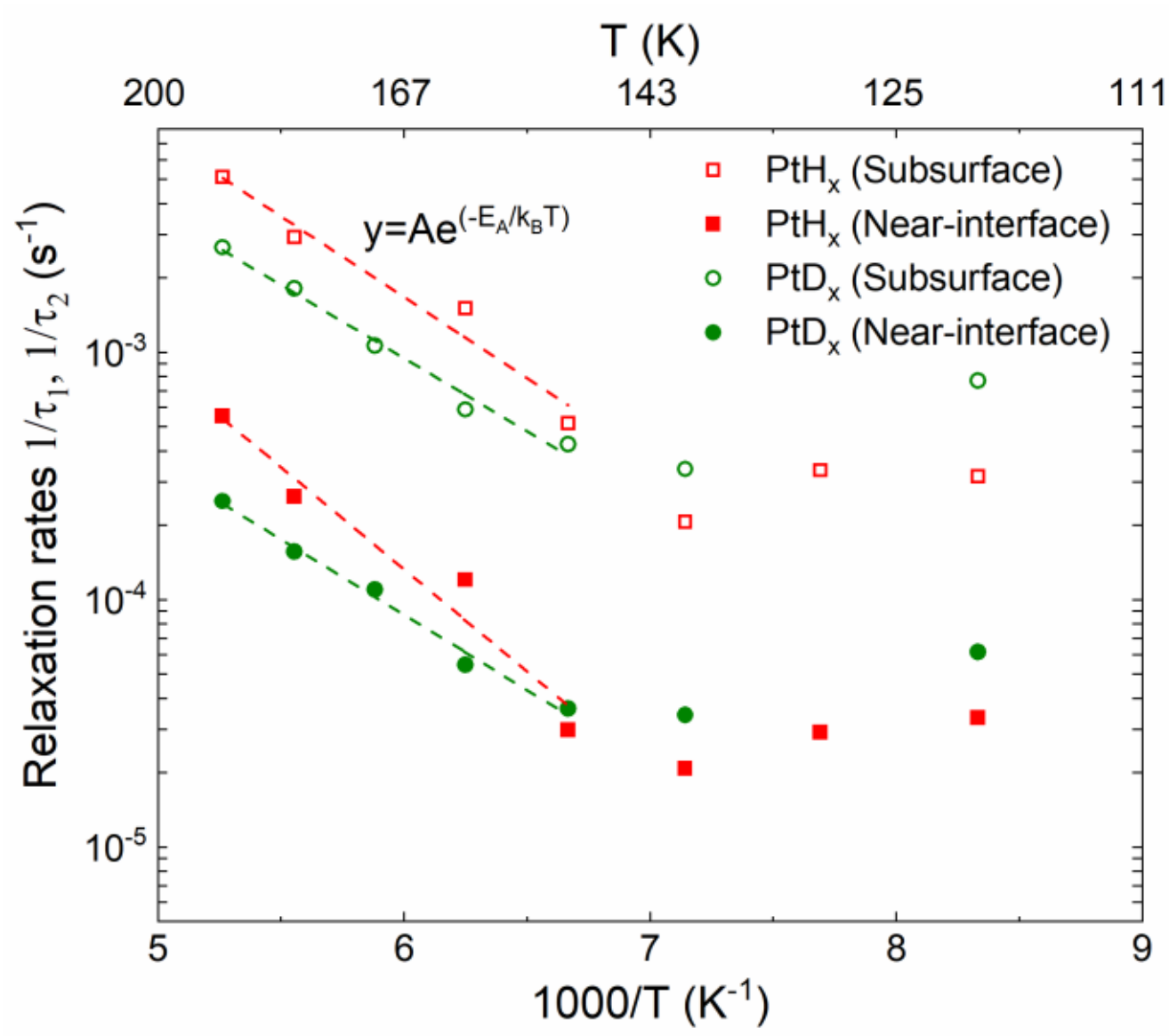

**Fig. 5** Arrhenius plot of the temperature dependence of the relaxation rates $1/\tau_1$ (subsurface component) and $1/\tau_2$ (near-interface component) for the $PtH_x$ film with $x_{Ave}$ = 0.15 and corresponding $PtD_x$ film prepared under the same implantation dose. The relaxation rates were obtained from fits to the resistance relaxation data using Eq. (3). $PtH_x$ and $PtD_x$ denote the H- and D-implanted Pt films, respectively. The subsurface component corresponds to the faster relaxation, whereas the near-interface component corresponds to the slower relaxation. The dashed lines represent Arrhenius fits in the thermally activated regime above 140 K. A clear isotope dependence is observed in this thermally activated regime (above 140 K), with deuterium exhibiting slower relaxation kinetics, whereas both isotopes show nearly identical relaxation behavior in the low-temperature regime.

**D. TDS Simulation Supporting Two Hydrogen Regions**

To support the two-region desorption kinetics, we extend the same kinetic framework to simulate the thermal desorption spectra for comparison with the experimental TDS data. The TDS was simulated by tracking the temperature-dependent evolution of hydrogen amount $\theta$ at a constant heating rate β (K/s). The desorption kinetics were modeled using the Polanyi-Wigner equation [30,31]

$$\frac{d\theta}{dT} = -\frac{\nu}{\beta}\,\theta^{n}\exp\left(\frac{-E_d}{k_B T}\right), \qquad (4)$$

where ν is the desorption prefactor, $E_d$ is the desorption activation energy, and $k_B$ is Boltzmann's constant. Although the Polanyi-Wigner equation is generally used for surface desorption, it can also be used as an effective phenomenological description for bulk or near-surface hydrogen release when the rate-limiting step is thermally activated detrapping and subsequent diffusion to the surface, as in the case of the present $PtH_x$ film. In the present model, the two hydrogen sites, subsurface and near-interface, were independently considered to contribute to the total TDS signal such that:

$$\theta = \theta_1 + \theta_2, \qquad (5)$$

$$\left(-\frac{d\theta}{dt}\right) = \left(-\frac{d\theta_1}{dt}\right) + \left(-\frac{d\theta_2}{dt}\right) \qquad (6)$$

Here, $\theta_1$ and $\theta_2$ correspond to the subsurface and near-interface hydrogen amount. Although the desorption order cannot be uniquely defined for bulk hydrogen release, as in the case of surfaces, the first-order approximation ($n$ = 1) is adopted as a minimal physically reasonable model, as it follows the experimental TDS shape and relative contributions from the subsurface and near-interface components. Figure 6 shows the simulated TDS spectrum for $PtH_x$ film with $x_{Ave}$ = 0.15, calculated using the kinetic parameters obtained from the Arrhenius analysis in Fig. 5. The corresponding simulation for the $PtD_x$ film prepared under the same implantation dose is shown in Fig. S7. Although the two desorption components are strongly overlapping at this lower hydrogen concentration, the simulation reveals that the experimentally observed broad TDS feature arises from the superposition of two distinct desorption processes. For the higher-concentration $PtH_x$ film with $x_{Ave}$ = 0.34, the tendency of the two desorption peaks was also reproduced, as shown in Fig. S8. The simulations were performed for the $PtH_x$ samples with $x_{Ave}$ = 0. 15 and 0.34 because these two concentrations represent respectively as lower- and higher-concentration regimes. For $x_{Ave}$ = 0.15, the two TDS components strongly overlap, whereas for $x_{Ave}$ = 0.34, the two desorption peaks become more clearly resolved. Thus, these two cases were chosen to verify whether the same kinetic framework can reproduce the evolution of the TDS spectra with increasing hydrogen concentration rather than fitting every implantation dose separately. These simulations establish a qualitative framework linking resistance relaxation, TDS and NRA results, and also support the interpretation that hydrogen in $PtH_x$ occupies subsurface and near-interface regions with distinct desorption and relaxation kinetics.

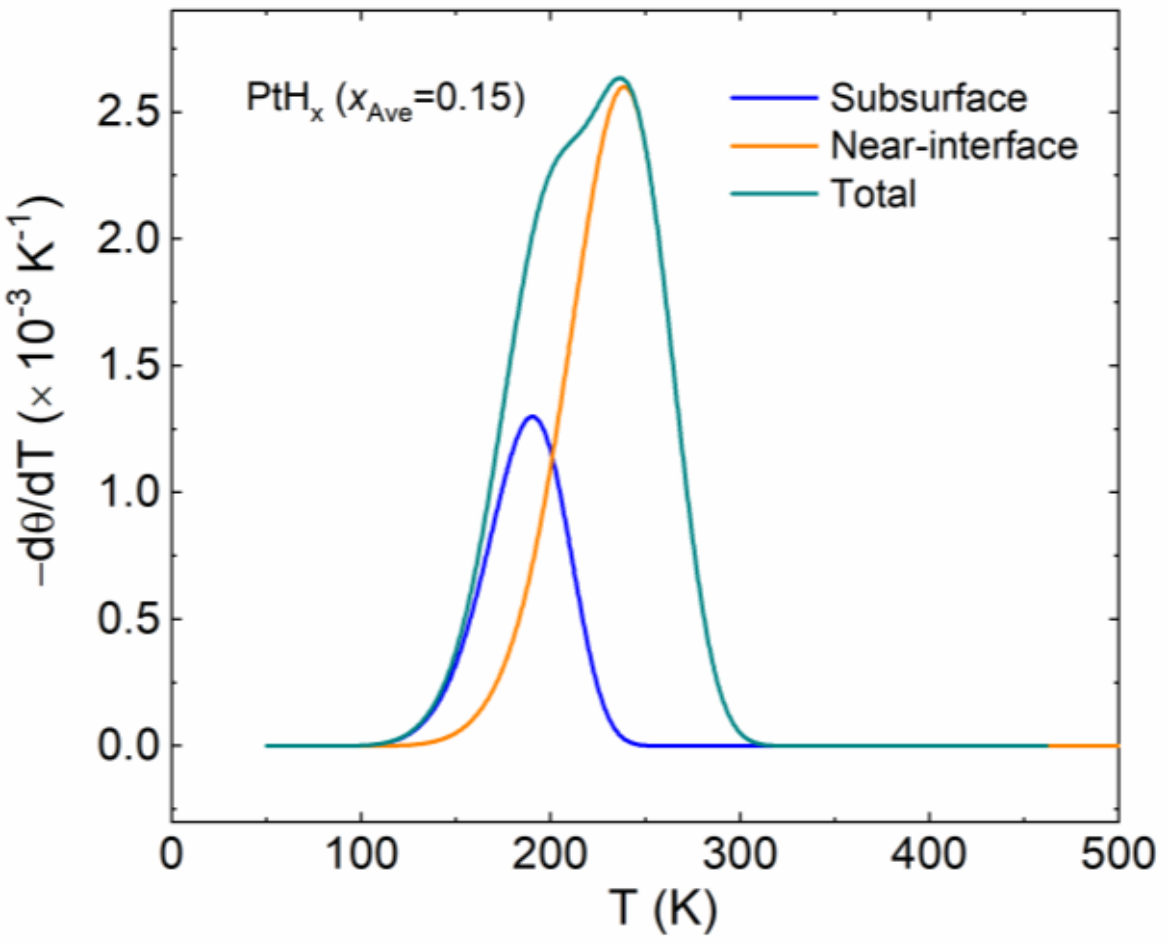

**Fig. 6** Simulated TDS data for $PtH_x$ film with $x_{Ave}$ = 0.15, displaying the subsurface and near-interface hydrogen contributions and their combined desorption signal. The desorption rate (-dθ/dT) is plotted as a function of temperature, where θ is the normalized hydrogen concentration. The simulation is performed using eq. (4) with the following initial parameters obtained from the hydrogen depth profile in Fig. 1 and the Arrhenius analysis in Fig.5 : $T_0$ = 50 K, β = 0.35 K/s, desorption order n = 1, $\theta_1$(subsurface) = 0.2, $E_d$(subsurface) = 0.13 eV, ν(subsurface) = 14.2 $s^{-1}$, $\theta_2$(near-interface) = 0.5, $E_d$(near-interface) = 0.164 eV, ν(near-interface) = 11.9 $s^{-1}$.

**Table 1.** Summary of the TDS peak temperatures and kinetic parameters obtained from the resistance relaxation analysis. The H data correspond to the $PtH_x$ sample prepared with an implantation dose of 4.7 mC cm$^{-2}$ and an average hydrogen concentration of $x_{Ave} = 0.15$. The D data correspond to the $PtD_x$ sample prepared under the same implantation dose.

| Isotope | TDS peak $T_1$ (K) | TDS peak $T_2$ (K) | $E_{a_1}$ (meV) | $E_{a_2}$ (meV) | $A_1$ ($s^{-1}$) | $A_2$ ($s^{-1}$) |
|---|---|---|---|---|---|---|
| **H** | ~190 | ~230 | 130 ± 18 | 164 ± 26 | 14.2 ± 3.6 | 11.9± 6.0 |
| **D** | ~180 | ~220 | 117 ± 8 | 121 ± 7 | 3.4 ± 1.8 | 0.39 ± 1.6 |

**E. Conclusion**

To conclude, we have presented a comprehensive investigation of hydrogen incorporation, depth distribution, desorption kinetics, and isotope effects in metastable $PtH(D)_x$ thin films prepared by low-energy ion irradiation at low temperature. NRA depth profiling reveals a nonuniform hydrogen depth distribution, with hydrogen accumulation in the subsurface and film-substrate near-interface regions. TDS data showed two overlapping desorption peaks, indicating hydrogen populations with different thermal stabilities and binding environments. The resistance relaxation dynamics of $PtH_x$ are governed by hydrogen desorption with two kinetically distinct components associated with hydrogen release from subsurface and near-interface regions. Based on NRA, TDS and resistance relaxation results, it is shown that the near-interface region is thermally more stable than the subsurface region. Arrhenius analysis of the resistance relaxation above 140 K shows that the relaxation for H is faster than that for D. Although the activation energies for H and D are comparable within uncertainty, the systematically slower relaxation of D indicates isotope-dependent kinetics likely influenced by reduced attempt frequencies and zero-point energy effects. In contrast, at low temperature (< 140 K), the relaxation rates for H and D become nearly temperature-independent regardless of the isotope. Overall, these findings provide insight into the synthesis of nonequilibrium hydrogen states and isotope-dependent hydrogen dynamics in Pt thin films, relevant to Pt-based catalysis, sensing, and related hydrogen-energy technologies.

**Acknowledgements:** The authors thank H. Matsuzaki, T. Yamagata, and H. Tokuyama at the Micro Analysis Laboratory, Tandem accelerator (MALT) facility at the University of Tokyo for providing beam time and technical support during the NRA measurements. This work was supported by JSPS KAKENHI Grant Numbers JP21H04650, JP25K24643, and JP24K17612, and by JST PRESTO (Grant Number JPMJPR2504), Japan.

**Hydrogen (deuterium) dynamics and thermal stability in ion-irradiated platinum-hydride thin films synthesized at low temperature**

S. S. Das[1,*] , T. Ozawa[1,*], Y. Komatsu[2], R. Shimizu[2,†], T. Hitosugi[2,3] , and K. Fukutani[1,4,*]

[1]*Institute of Industrial Science, The University of Tokyo, Komaba, Meguro-ku, Tokyo, 153-8505, Japan*

[2]*School of Materials and Chemical Technology, Institute of Science Tokyo, Ookayama, Meguro, Tokyo 152-8552, Japan.*

[3]*Department of Chemistry, The University of Tokyo, Hongo, Bunkyo, Tokyo 113-0033, Japan.*

[4]*Advanced Science Research Center, Japan Atomic Energy Agency, Shirakata, Tokai, Ibaraki 319-1195, Japan.*

*Present address:*

[†]*Institute for Molecular Science, National Institutes of Natural Science, Myodaiji, Okazaki, Aichi 444-8585, Japan.*

## Supplementary file

[*] Corresponding authors: ssdas@iis.u-tokyo.ac.jp (S. S. Das), t-ozawa@iis.u-tokyo.ac.jp (T. Ozawa), and fukutani@iis.u-tokyo.ac.jp (K. Fukutani)

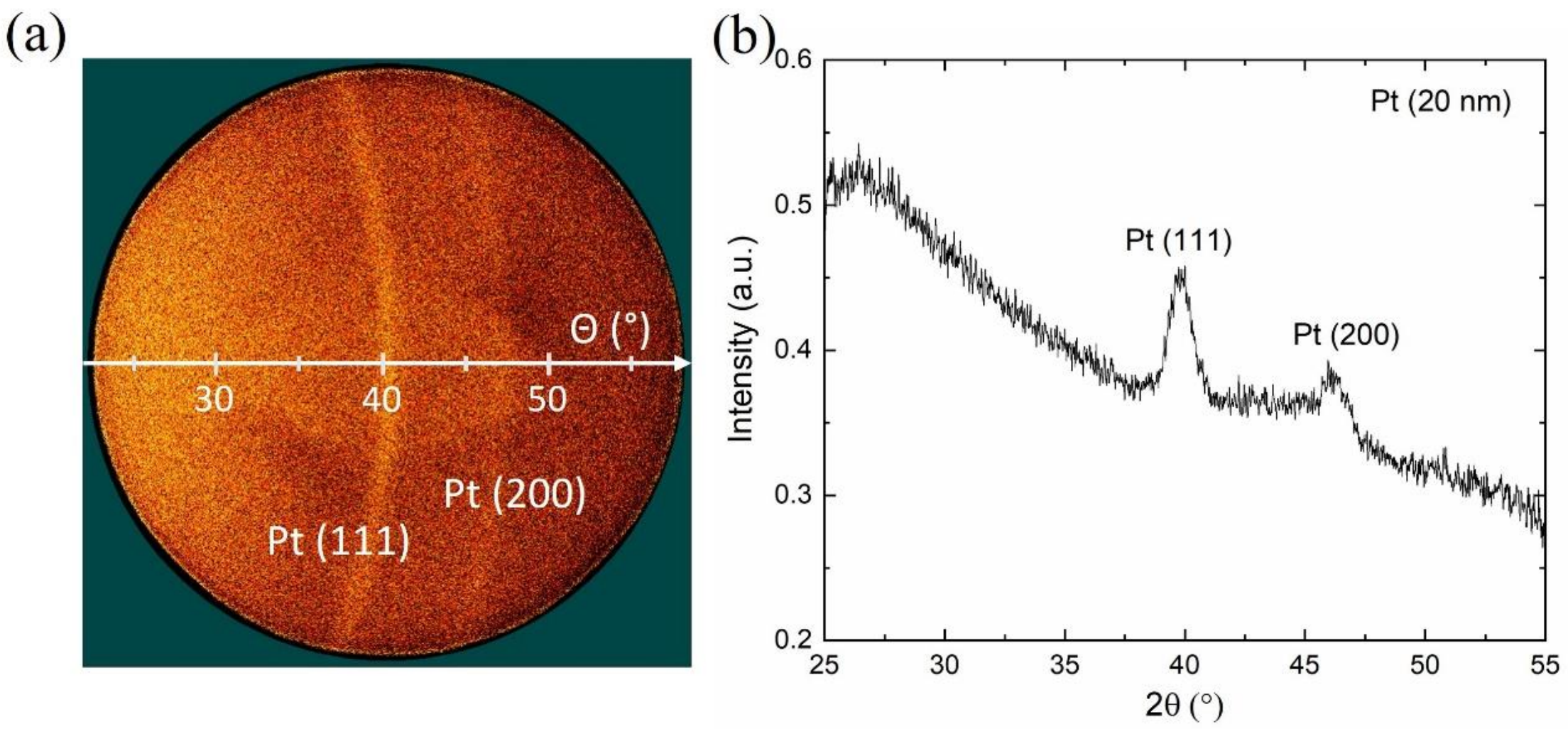

**Fig. S1** Structural characterization of the Pt(20 nm) film via 2D X-ray Diffraction. (a) 2D-XRD profile exhibiting characteristic Debye-Scherrer diffraction rings. The continuous nature of the rings suggests a polycrystalline structure with a preferred texture. (b) Integrated 1D-XRD intensity profile as a function of 2θ. The diffraction peaks at approximately 40° and 46° correspond to the (111) and (200) planes of Pt, respectively. The broad background at lower angles (centered around 25°) is attributed to the amorphous substrate or sample holder.

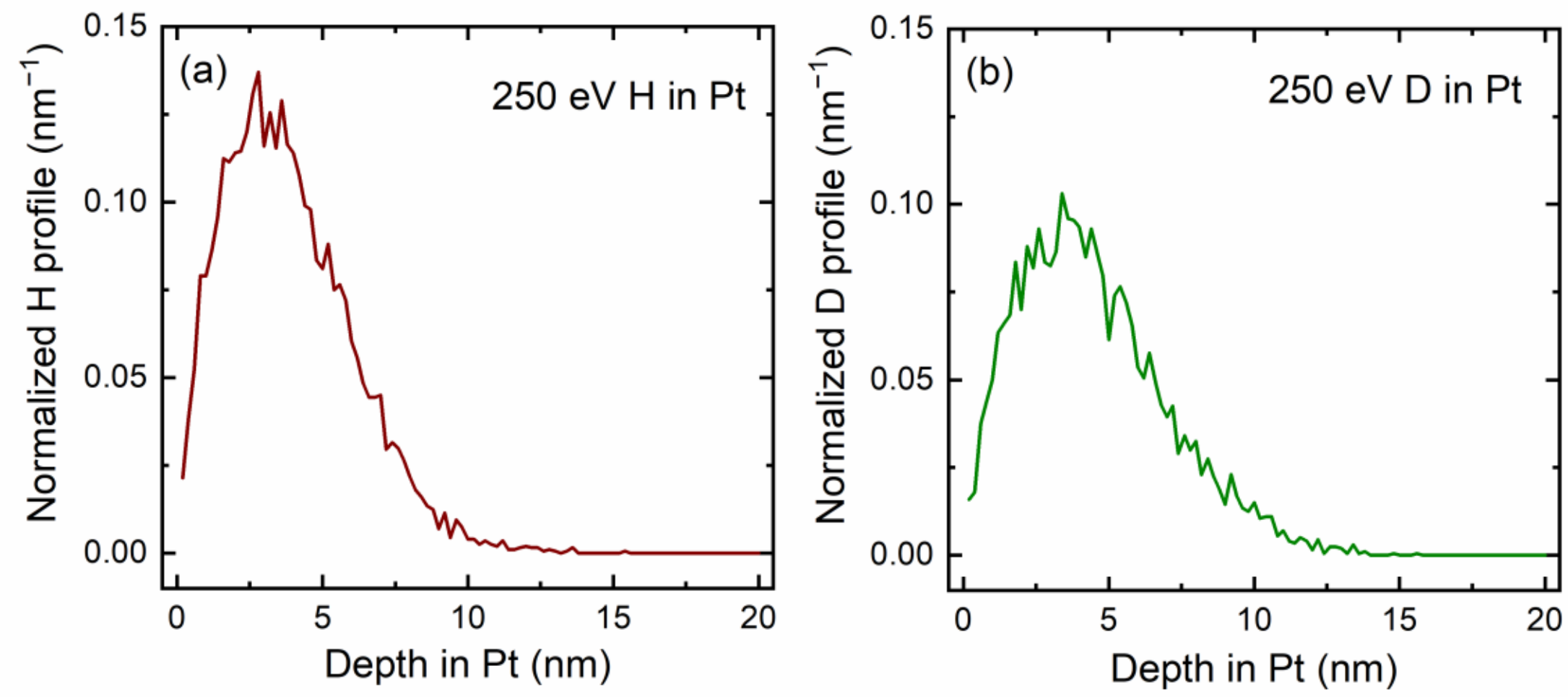

**Fig. S2** Stopping and Range of Ions in Matter (SRIM) simulated implantation profiles of 250 eV (a) H and (b) D ions in a 20 nm Pt film. The simulations suggest shallow near-surface accumulation for both isotopes, with projected ranges of 3.7 nm for H and 4.4 nm for D. The corresponding longitudinal straggles are ~2.2 nm for H and ~2.4 nm for D.

**Estimation of hydrogen retention fraction**

The nominal implanted hydrogen amount was estimated from the implanted charge density. In the implantation process, the implanted ion beam mainly consisted of molecular $H_2^+/D_2^+$ ions. The number of $H_2^+$ ions per dose of $H_{imp}$ [mC/cm²] corresponds to:

$N(H_2^+) = 10^{-3} / (1.602 \times 10^{-19}) \times H_{imp}$ [ions/cm²]

Since each $H_2^+$ ion contains two H atoms, the number of H atoms is $N(H) = 2 \times N(H_2^+)$.

The number of Pt atoms per unit area in a film with a thickness of $d$ is $nd$ with the atomic density ($n$) for Pt ($6.63 \times 10^{22}$ atoms/cm³). Then, the nominal H concentration is described as:

$x^n = N(H)/nd.$

The retention fraction was estimated as $x_{Ave} / x^n$ using the average concentration ($x_{Ave}$) derived from the integrated NRA profile.

**Table-S1:** Estimated hydrogen retention fraction calculated from the ratio of the NRA-derived average hydrogen concentration $x_{Ave}$ to the nominal implanted hydrogen concentration $x^n$.

| Implantation dose (mC/cm²) | $x^n$ (H/Pt) | $x_{Ave}$ (H/Pt) | Retention fraction ($x_{Ave}/x^n$) |
|---|---|---|---|
| 1.56 | 0.147 | 0.089 | 0.61 |
| 4.69 | 0.442 | 0.155 | 0.35 |
| 9.38 | 0.884 | 0.245 | 0.28 |
| 14.06 | 1.325 | 0.326 | 0.25 |
| 18.75 | 1.768 | 0.369 | 0.21 |
| 28.13 | 2.652 | 0.415 | 0.16 |
| 39.06 | 3.682 | 0.422 | 0.11 |
| 78.13 | 7.365 | 0.403 | 0.05 |
| 156.25 | 14.729 | 0.410 | 0.03 |

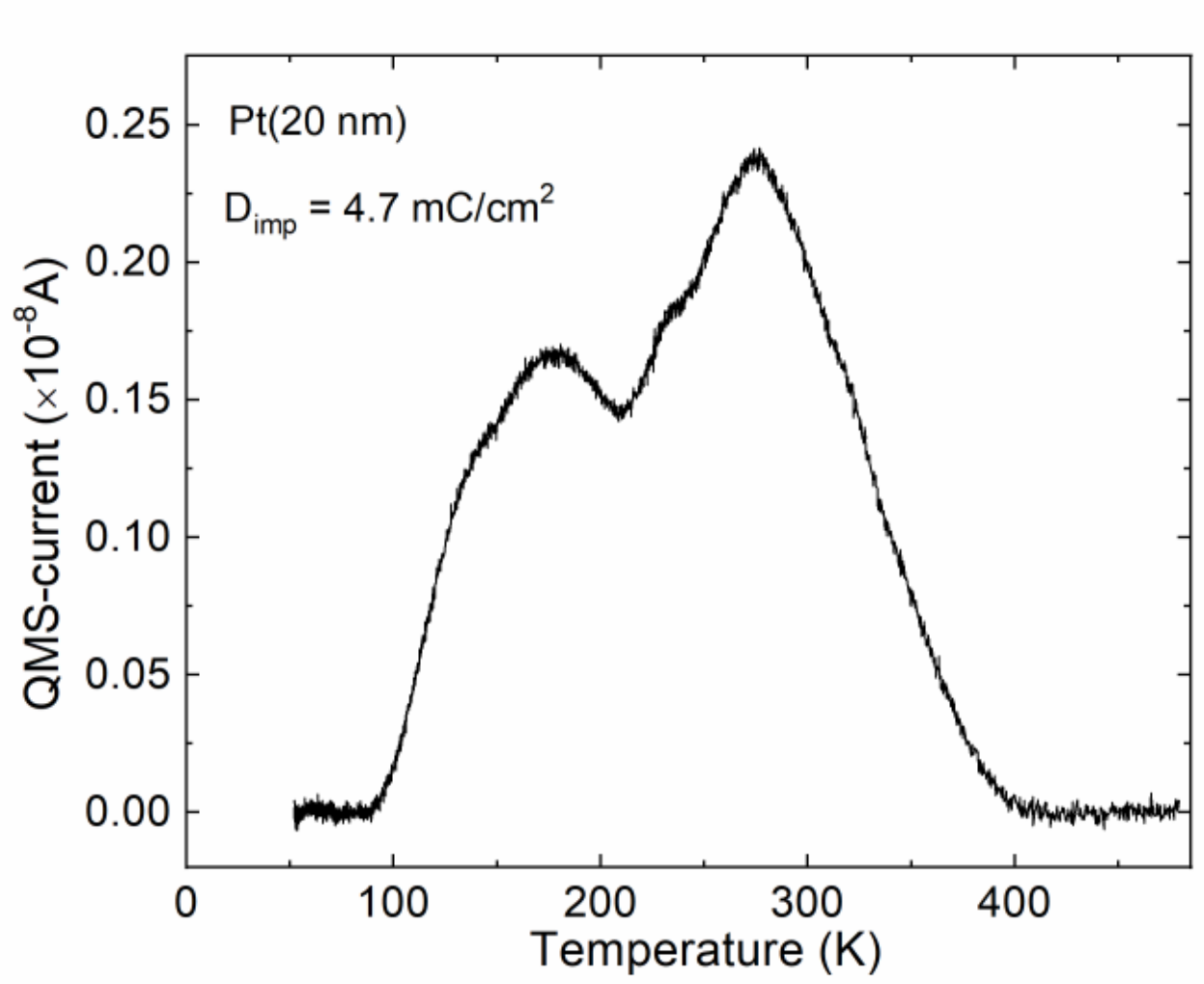

**Fig. S3** Thermal desorption spectroscopy (TDS) spectrum of Pt(20 nm) film following deuterium implantation with a dose $D_{imp}$ = 4.7 mC/cm$^2$. The QMS signal for $D_2$ is plotted as a function of temperature, showing a broad spectrum consisting of two partially overlapping peaks, indicative of deuterium release from subsurface and near-interface regions in the $PtD_x$ film.

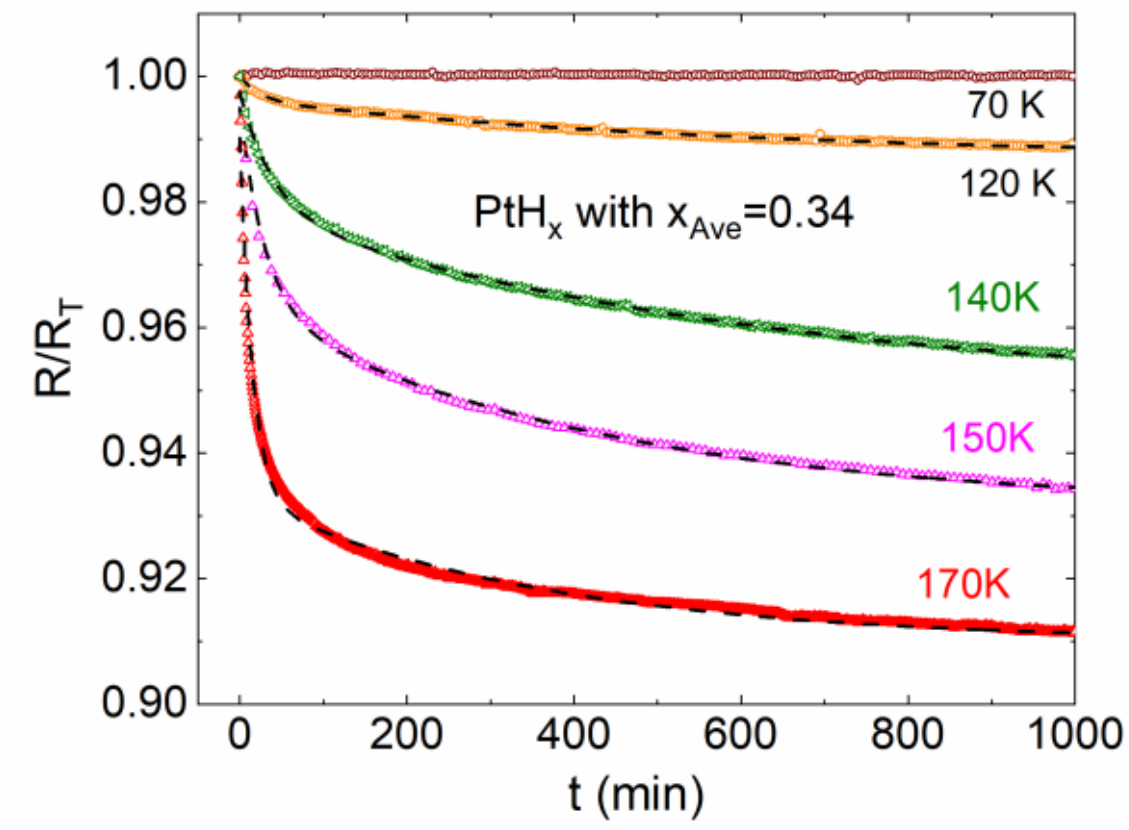

**Fig. S4** Time evolution of the normalized resistance $R/R_T$ for $PtH_x$ film with $x_{Ave}$ = 0.34 measured at fixed temperatures of 70, 120, 140, 150, and 170 K. Here, $R_T$ denotes the resistance at the measurement temperature immediately before the onset of relaxation. The relaxation is negligible at 70 K, becomes weak at 120 K, and increases progressively with temperature above

140 K. The dashed lines represent fits to the experimental data using Eq. (3), highlighting the temperature-dependent relaxation dynamics.

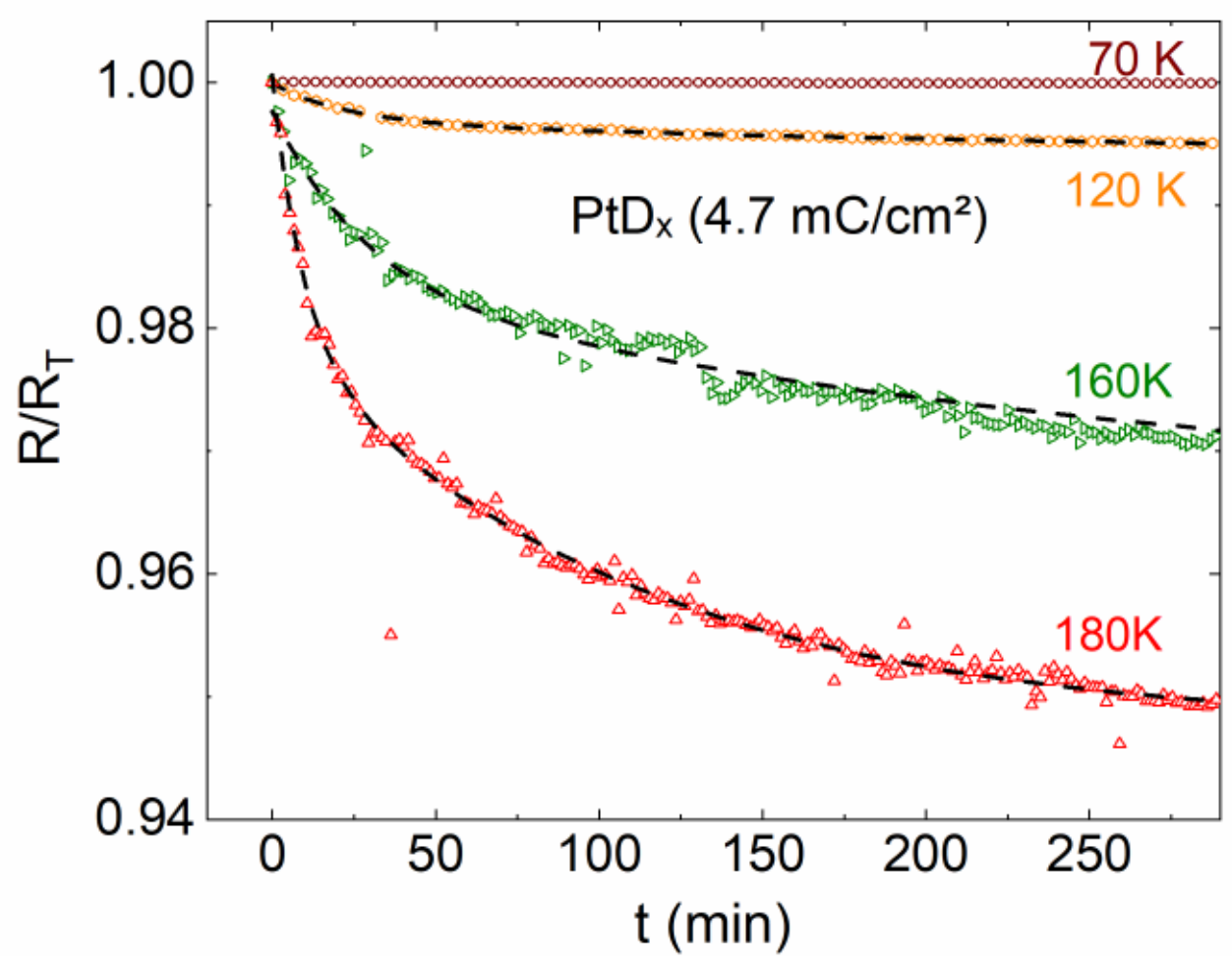

**Fig. S5** Time evolution of the normalized resistance $R/R_T$ for $PtD_x$ film measured at fixed temperatures of 70, 120, 160, and 180 K following deuterium implantation. The $PtD_x$ film was prepared using the same implantation dose as the $PtH_x$ film with $x_{Ave} = 0.15$. Here, $R_T$ denotes the resistance at the measurement temperature immediately before the onset of relaxation. The relaxation is negligible at 70 K, becomes weak at 120 K, and accelerates progressively at higher temperatures, reflecting thermally activated hydrogen release. The dashed lines represent fits to the experimental data using the two-channel model using eq (3) of main text.

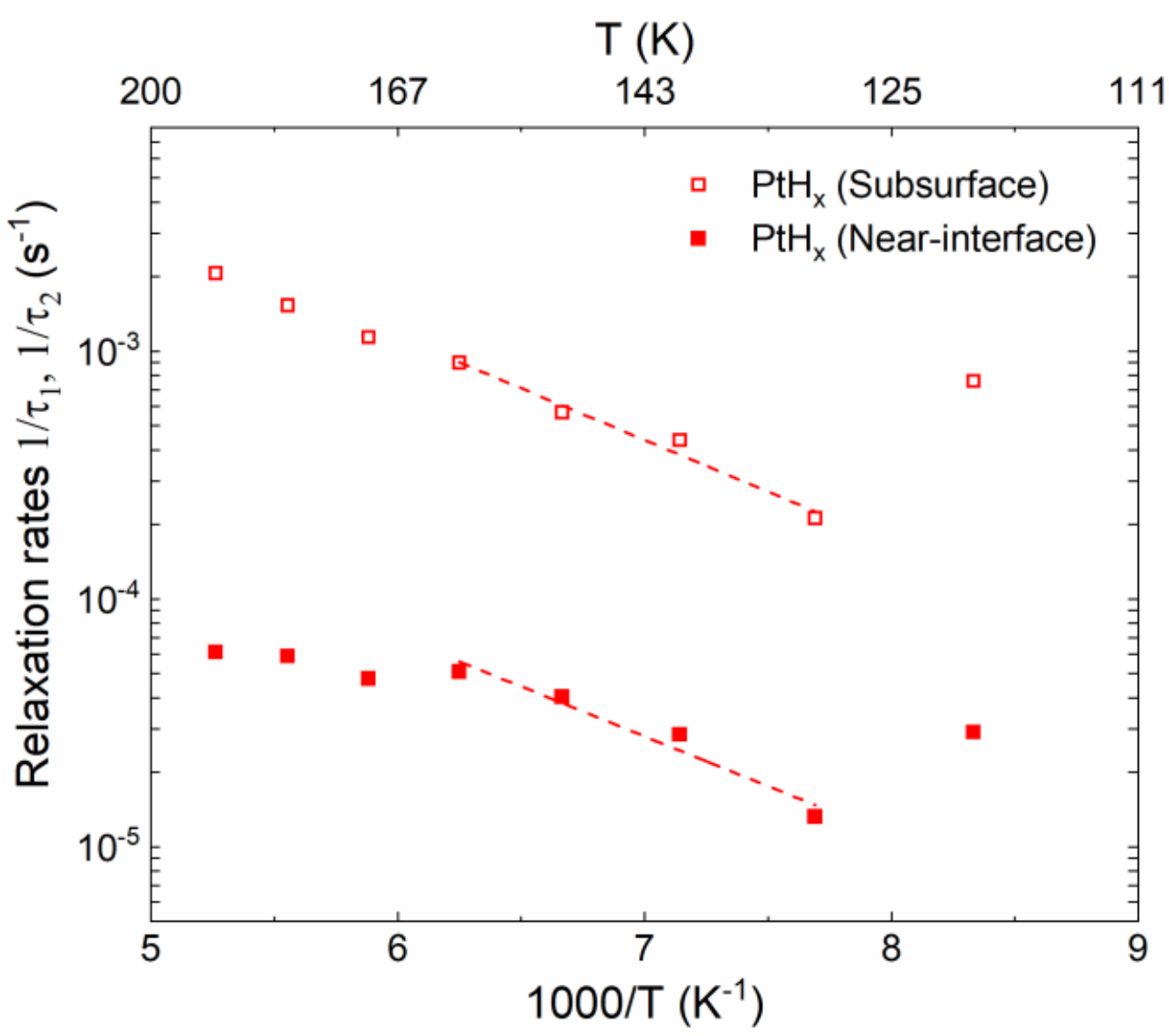

**Fig. S6** Arrhenius plot of the relaxation rates $1/\tau_1$ (subsurface, open symbols) and $1/\tau_2$ (near-interface, solid symbols) extracted from resistance relaxation measurements for $PtH_x$ film with $x_{Ave} = 0.34$. The dashed lines represent Arrhenius fits in the thermally activated regime, giving activation energy of $E_a \sim 83 \pm 9$ meV for the subsurface and $80 \pm 12$ meV for the near-interface components. Similarly, the corresponding prefactors are 0.37 and 0.018 $s^{-1}$. The Arrhenius parameters were extracted from the linear regime between 130–160 K; data above 160 K, where deviations from thermally activated behavior appear, were excluded from the fit.

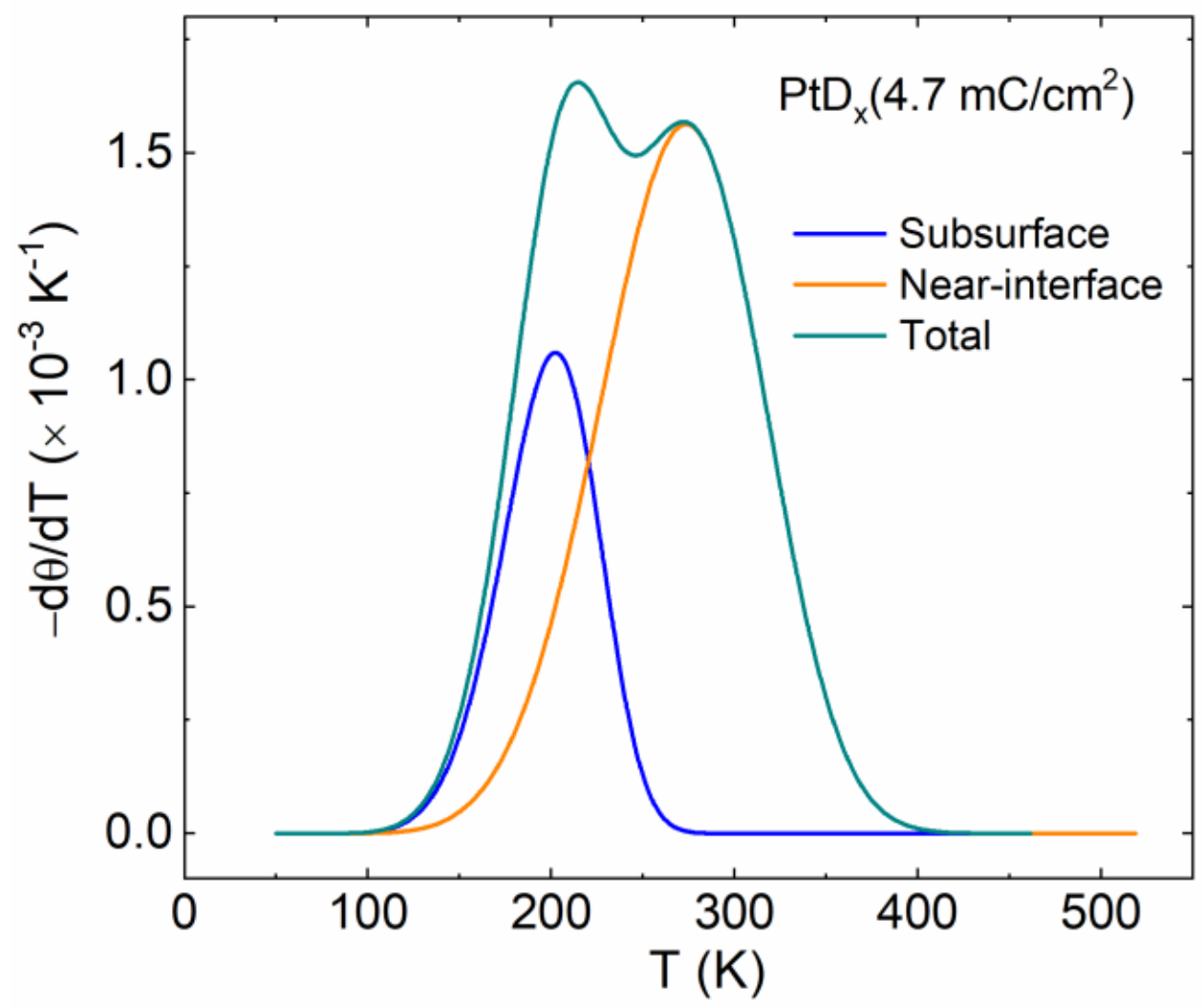

**Fig. S7** Simulated thermal desorption spectrum for $PtD_x$ film prepared with an implantation dose of 4.7 mC/cm$^2$, showing the individual contributions from subsurface (blue) and near-interface (yellow) deuterium populations, along with their combined desorption signal (total, green). The desorption rate ($-d\theta/dT$) is plotted as a function of temperature, where $\theta$ denotes the normalized deuterium amount The presence of two overlapping desorption components reflects kinetically distinct deuterium populations associated with subsurface and near-interface regions. The simulation is performed using eq. (4) with the following parameters: $T_0$ = 50 K, $\beta$ = 0.35 K/s, desorption order n = 1, $\theta_1$(subsurface) = 0.2, $E_d$(subsurface) = 0.117 eV, $\nu$(subsurface) = 3.35 s$^{-1}$, $\theta_2$(near-interface) = 0.5, $E_d$(near-interface) = 0.121 eV, $\nu$(near-interface) = 0.39 s$^{-1}$.

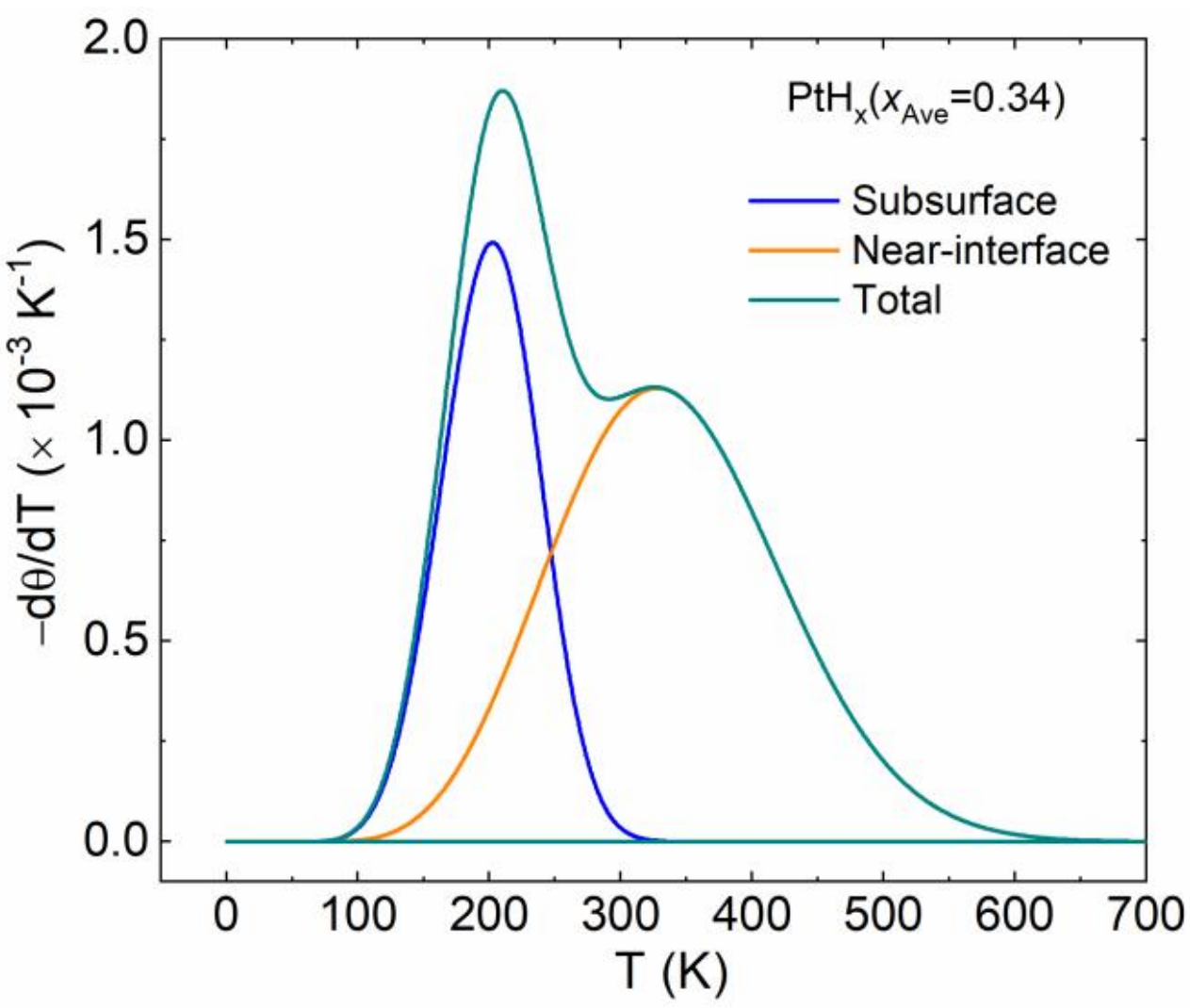

**Fig. S8** Simulated TDS data for $PtH_x$ film with $x_{Ave}$ = 0.34, showing the individual contributions from the subsurface (blue) and near-interface (yellow) hydrogen populations, along with their combined desorption signal (green). The desorption rate is plotted as a function of temperature. The two distinct desorption components reflect hydrogen release from subsurface and near-interface regions. The simulation is performed using eq. (4) with the following parameters: $T_0$ = 50 K, β = 0.35 K/s, desorption order n = 1, $\theta_1$(subsurface) = 0.4, $E_d$(subsurface) = 0.083 eV, ν(subsurface) = 0.037 $s^{-1}$, $\theta_2$(near-interface) = 0.7, $E_d$(near-interface) = 0.080 eV, ν(near-interface) = 0.018 $s^{-1}$.